\documentclass[conference,a4paper]{IEEEtran}

\usepackage{etex}
\usepackage{mathtools}
\usepackage{amssymb}
\usepackage{amsmath}
\usepackage{algorithm,algorithmicx,algpseudocode}

\usepackage{pictex,graphicx}

\usepackage{cite}

\usepackage{array}

\usepackage[tight,footnotesize]{subfigure}

\newtheorem{remark}{Remark}

\begin{document}

\sloppy

\title{A Deterministic Annealing Approach to Optimization of Zero-delay Source-Channel Codes}

\author{
  \IEEEauthorblockN{Mustafa S. Mehmetoglu, Emrah~Akyol, Kenneth Rose}
  \IEEEauthorblockA{Dep. of Electrical \& Computer Eng.\\
    UC Santa Barbara, CA, US\\
    Email: \{mehmetoglu, eakyol, rose\}@ece.ucsb.edu} 
}



\maketitle

\begin{abstract}
This paper studies optimization of zero-delay source-channel codes, and specifically the problem of obtaining globally optimal  transformations that  map between the source space and the channel space, under a given transmission power constraint and for the mean square error distortion. Particularly, we focus on the setting where the decoder has access to side information, whose cost surface is known to be riddled with local minima. Prior work derived the necessary conditions for optimality of the encoder and decoder mappings, along with a greedy optimization algorithm that imposes these conditions iteratively, in conjunction with the heuristic ``noisy channel relaxation" method to mitigate poor local minima. While noisy channel relaxation is arguably effective in simple settings, it fails to provide accurate global optimization results in more complicated settings including the decoder with side information as considered in this paper. We propose a global optimization algorithm based on the ideas of ``deterministic annealing"- a non-convex optimization method, derived from information theoretic principles with analogies to statistical physics, and successfully employed in several problems including clustering, vector quantization and regression. We present comparative numerical results that show strict superiority of the proposed algorithm over greedy optimization methods as well as over the noisy channel relaxation.
\end{abstract}

\section{Introduction}
The zero delay source-channel coding problem has recently gained revived interest \cite{hekland2009shannon,hu2011analog,vaishampayan2003curves,chen2011zero,karlsson2010optimized}. In this paper, we focus on numerical optimization of the zero-delay mappings. In prior work\cite{ emrah_dcc10}, a method, ``noisy channel relaxation" (NCR) \cite{gadkari1999robust, Knagenhjelm} was employed to mitigate the poor local minima problem inherent to such optimization problems. While NCR is relatively successful in the point-to-point setting, it is insufficient to obtain precise results in more involved settings such as the decoder side information setting. In this paper, we incorporate a powerful non-convex optimization method, {\it deterministic annealing},  within a framework proposed in our prior work  \cite{emrah_dcc10} to numerically obtain the globally optimal zero-delay mappings in the side information setting. 

Deterministic annealing (DA) is a global optimization approach, based on information theoretic principles with analogies to statistical physics, that has been successfully used as a remedy to the problem of poor local minima in non-convex optimization problems, including clustering \cite{rose1990statistical}, vector quantization \cite{rose1992vector}, regression \cite{rao1999deterministic} and more (see review in \cite{da}). An important distinction between DA and other non-convex optimization tools such as NCR is that DA is independent of the initialization.
 
This paper is organized as follows. In Section II, we present preliminaries and the problem definition. In Section III, we review prior work including the necessary conditions for optimality, and optimization aided by NCR. In Section IV, we describe the proposed algorithm. Numerical comparisons are presented in Section V and  concluding remarks in Section VI.

  \begin{figure}
\centering \includegraphics[width=1 \linewidth]{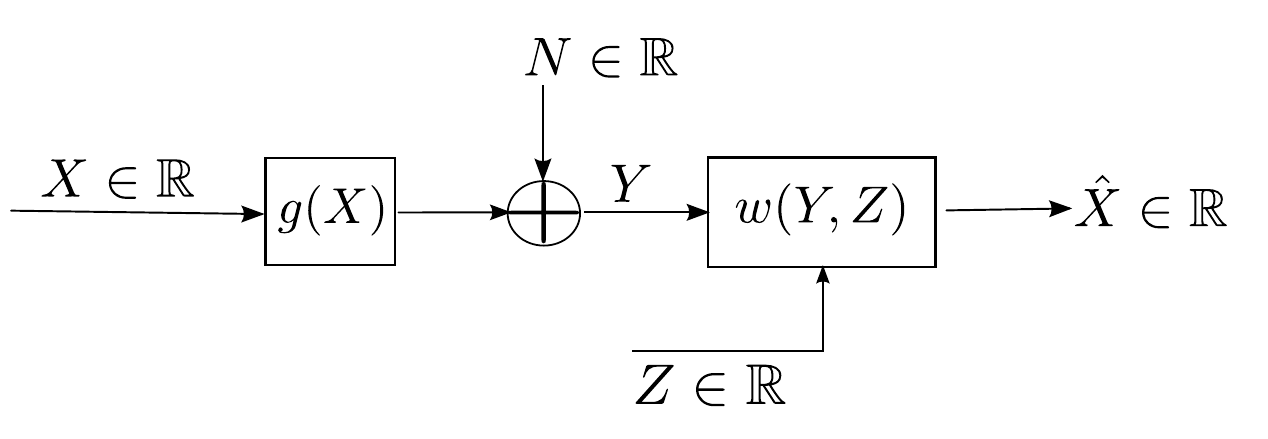}
\caption{The problem setting}
\label{side} 
\end{figure} 

\section {Preliminaries and Problem Definition}
Let $\mathbb E(\cdot)$, $\mathbb P(\cdot)$ and $\mathbb R$ denote the expectation and probability operators, and the set of real numbers, respectively. Let $\nabla$ and $\nabla_x$ denote the gradient and partial gradient with respect to $x$, respectively. Let $f^{'} (x)= \frac{d f(x)}{d x}$ denote the first order derivative of function $f(\cdot)$. The joint Gaussian density with mean $\boldsymbol \mu$ and covariance matrix $R$ is denoted as $\mathcal N( \boldsymbol \mu,R)$. All the logarithms in the paper are natural logarithms and may in general be complex. The integrals are in general Lebesgue integrals. While we focus on scalar sources and noises, our results can easily be extended to vector spaces, albeit with more involved notations. 

The problem setting is given in Figure \ref{side}, where source ${X} \in  \mathbb R $ and side information ${ Z} \in \mathbb R $ are drawn from joint density $f_{X,Z}( \cdot,\cdot)$.  $Z$ is available only to the decoder, while $ X$ is mapped to channel input by the encoding function ${ g}:\mathbb R\rightarrow \mathbb R$ and transmitted over the channel whose additive noise ${ N} \in \mathbb R$, with density $f_N(\cdot)$, is independent of $ X,  Z$. The received channel output ${  Y}=g(X)+N$ is mapped to the estimate ${\hat X}$ by the decoding function ${ w} :  \mathbb R \times \mathbb R  \rightarrow  \mathbb R$. The problem is to find optimal mapping functions $ g (\cdot), w(\cdot)$ that minimize the mean squared error (MSE) distortion
\begin{equation}
 D=\mathbb E\{(X-\hat X)^2\},
 \label{distortion1}
 \end{equation}
subject to 
  \begin{equation}
 P(g)=\mathbb E\{ g^2(X) \} \leq P.
 \label{power_cons1}
 \end{equation}

 Although the problem we consider is delay limited, it is insightful to consider asymptotic bounds achievable at infinite delay. From Shannon's source and channel coding theorems, it is known that, asymptotically, the source can be compressed to $R(D)$ bits (per source sample) at distortion level $D$, and that $C$ bits can be transmitted over the channel (per channel use) with arbitrarily low probability of error, where $R(D)$ is the source rate-distortion function, and $C$ is the channel capacity, (see e.g.\cite{coverbook}). The asymptotically optimal coding scheme is the tandem combination of the optimal source and channel coding schemes,  hence $ R(D) \leq C$ must hold. By setting 
\begin{equation}
\label{eq1}
R(D) =C,
\end{equation}
one obtains a lower bound on the distortion of any source-channel coding scheme. The capacity of the additive white gaussian noise channel with variance $\sigma_N^2$ is given by 
\begin{equation}
\label{eq3}
C=\frac{1}{2}\log(1+\frac{P}{\sigma_N^2}),
\end{equation}
where $P$ is the transmission power constraint and $\sigma_N^2$ is the noise variance. 
For source coding with decoder side information, it has been established for Gaussians and MSE distortion that there is no rate loss due to the fact that the side information is unavailable to the encoder \cite{wynerziv}. Hence, optimum performance theoretically attainable (OPTA) can be obtained by equating the conditional rate distortion function of the source (given the side information) to the channel capacity. The rate distortion function of $X$ when $Z$ serves as side information and $[X,Z] \sim \mathcal N(\boldsymbol 0, R)$ where $R=\sigma_X^2 \left[ \begin{array}{cc}  1 & \rho \\ \rho &1 \end{array} \right ]$ with $|\rho|\leq 1$ is:
\begin{equation}
\label{eq22}
R(D)=\max(0,\frac{1}{2}\log\frac {(1-\rho^2)\sigma_X^2}{D}),
\end{equation}
We plug (\ref{eq22}) and (\ref{eq3}) in (\ref{eq1}) to obtain OPTA
\begin{equation}
\label{opta2}
D_{OPTA}=\frac{(1-\rho^2)\sigma_X^2}{(1+\frac{P}{\sigma_N^2})}.
\end{equation}

 
\section {Prior Work}
Here, we summarize the relevant contributions of prior work, see \cite{emrah_dcc10} for more details. 
\subsection{Necessary Conditions for Optimality}
\label{nec}
Let the encoder $ g(\cdot)$ be fixed. Then, the optimal decoder is the MSE estimator  of $X$ given $Z=z$ and $Y=y$:
\begin{equation}
w({  y}, z)= \mathbb E \{ X| { y}, z\}.
\end{equation}
Plugging the expressions for expectation, applying Bayes' rule and noting that $f_{ Y|X}({ {  y},  x})= f_{N}[{ {  y}- g( x)}]$, the optimal decoder can be written, in terms of known quantities, as    
\begin{equation} \label{decoder_map}
{ w( { y}, z)}= \dfrac{ \int { x}  \,  f_{X,Z}( { x, z})  \,  f_{N} [y- g( x)]\, { \mathrm{d} x} } {  \int  \,      f_{X,Z}( { x,  z})  \,  f_{N} [{ { y}- g( x)}]  \,  \mathrm{d} x}.
\end{equation}
To derive the necessary condition for optimality of $ g(\cdot)$, we consider the distortion functional   
\begin{equation}
D[{ g, h}] =\mathbb E \{ ( {  X-  w( g( X)+N, Z)} )^2\},
\end{equation}
and construct the Lagrangian cost functional:
\begin{equation}\label{total_cost_SI}
J [{ g, w}] =D[{ g, w}]+\lambda P [{ g}] .
  \end{equation}
Now, let us assume the decoder $ w(\cdot)$ is fixed. To obtain necessary conditions, we apply the standard method in variational calculus:
\begin{equation}\label {encoder_map}
 \nabla_{ g} J [{ g}, w]=0, \, \, \forall \, x,
\end{equation} 
where 
 \begin{align}
 \nabla_{ g } J[ g,w] \!&= \lambda f_{X}(x)   { g( x)} \nonumber \\\! -\!\!\!\int \! \!\!\! \int   \! \! w'( g( x)\!+\! n, z)  
&   \left [{ x\!-\! w( g( x)\!+\! n, z)}\right]  f_N({ n}) f_{X,Z} ({  x, z})   \mathrm{d} n \mathrm{d}z.
\end{align}
and $w'(\cdot,\cdot)$ denotes the derivative with respect to the first argument. 
\begin{remark}
\label{linear necessary}
Note that the linear encoder and decoder mappings satisfy the necessary conditions for optimality in the Gaussian case.  However,  it is well known that linear mappings are highly suboptimal, see e.g. \cite{emrah_dcc10}. This fact  illustrates the existence of poor local optima and the challenges facing algorithms based on these necessary conditions.
\end{remark}

\subsection{Greedy Algorithm}
 Iteratively alternating between the imposition of individual necessary conditions for optimality, will successively decrease the Lagrangian cost to a stationary point. Imposing the decoder optimality condition is straightforward, since it is expressed in  closed form as functional of the encoding mapping ${ g} (\cdot)$. The encoder optimality condition is not in closed form and we perform an appropriate steepest descent search. The encoder is updated as given below, where $i$ is the iteration index and $\mu$ is the step size.
\begin{equation}
\label{g_iter} { g}_{i+1}({ x})= { g}_{i}({ x})-\mu \nabla_g {J[g,w]}.
\end{equation} 
At each iteration $i$, total cost decreases monotonically and iterations are kept until convergence. 

There is no guarantee that an iterative descent algorithm of this type will converge to the globally optimal solution, in fact, simulations show severe issues of local optima. As a  remedy,  NCR method of \cite{gadkari1999robust, Knagenhjelm} was embedded in the iterative algorithm in \cite{emrah_dcc10}, i.e., the algorithm was run for a very noisy channel (high Lagrangian parameter $\lambda$), and then gradually decrease $\lambda$ while using the prior mapping solution as initial condition. 

\section {Proposed Method}
We  recast the zero-delay source-channel coding problem as a regression problem optimizing for the encoding function  within a given parametric class of functions. We restrict the discussion to piecewise regression functions which approximate the desired mappings by partitioning the space and
matching a simple local model to each region. Such regression functions are determined by specifying two components: a  space partition and a parametric local model per partition cell (typically a simple model such as constant, linear, or Gaussian\footnote{In this paper, we use only affine models, however it is straightforward to include other models within the optimization framework.}).

DA introduces controlled randomization into the optimization process. The problem is recast as minimization of the expected cost subject to a constraint on the level of randomness as measured by the Shannon entropy of the system. The resulting Lagrangian functional can be viewed as the free energy of a corresponding physical system whose Lagrange parameter is the ``temperature". The minimization is started at a high temperature (highly random mappings) where, in fact, the entropy is maximized and all points equally belong to the all partition cells (and effectively there is only one local model). This minimum is then tracked at successively lower temperatures (lower levels of entropy) as the system typically undergoes a sequence of phase transitions through which the model complexity (the number of distinct local models) grows. As the temperature approaches zero, the distortion and power terms  dominate the Lagrangian cost and a hard (nonrandom) mapping is obtained.

We proceed to describe in more detail the proposed DA-based method.

\subsection {Structured Encoder Functions}
\begin{figure}
\includegraphics[width=1 \linewidth]{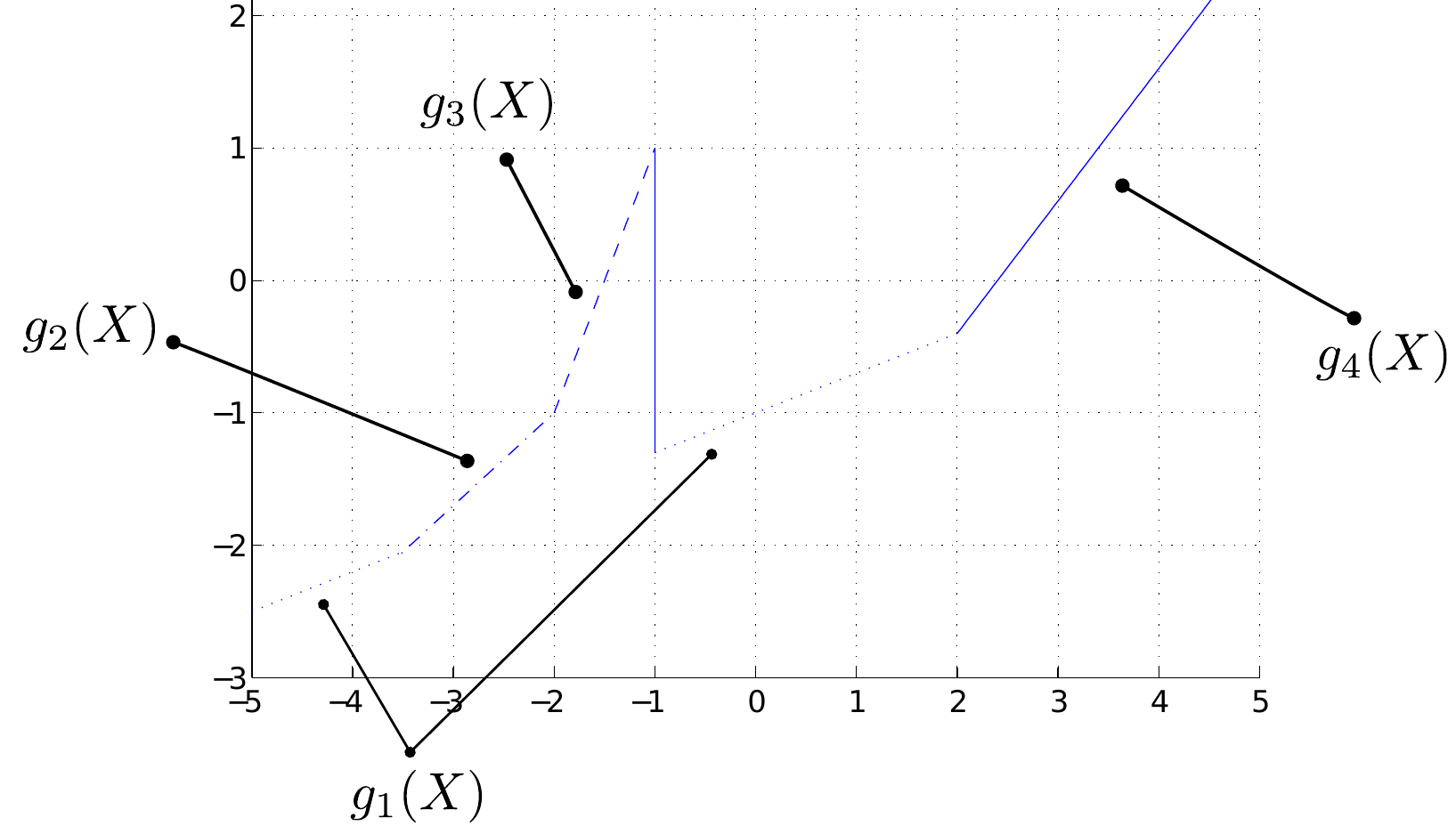}
\centering \caption{An example encoder function consisting of affine local models, $K=4$ }
\label{structuredEx} 
\end{figure}

We consider the parametric functions (local models) $g_k(x)=f(x,\Lambda_k)$, for $k\in \{1,...,K\}$, with the parameter sets $\Lambda_k$. These functions have a certain parametric form and each function is defined over a region denoted as $\mathbb R_k$. The overall encoder function is defined as $g(x)=g_k(x)$ for $x\in\mathbb R_k$.  The parametric form is to be chosen appropriately depending on the involved distributions and the design constraints. Figure \ref{structuredEx} shows an example structured encoder with affine local models of the form $g_k(x)=a_kx+b_k$.

\subsection{Randomized Associations}
We randomize the associations of the input points to the local models, or regions. We first define the probabilities
\begin{equation}
p_{K|X}(k|x)\triangleq \mathbb P\{ x \in \mathbb R_k\} , \quad \forall k,x.
\end{equation}
Note that $\sum\limits_{k=1}^{K} p_{K|X}(k|x) = 1$ $\forall x$. Next, we  rewrite (\ref{distortion1}) as
\begin{equation}
 D=\sum\limits_{k=1}^{K} \int\limits _{\mathbb R} D_k(x)p_{X}(x)p_{K|X}(k|x) \mathrm{d} x,
 \label{distortion_new1}
 \end{equation}
 where $D_k(x)$ is the contribution to the distortion, when point $x$ is associated with region $k$. It is given by 
 \begin{equation}
 D_k(x) = \int\limits _{\mathbb R} d(x,w(g_k(x)+n,z) )p_N(n)p_{Z|X}(z|x) \mathrm{d} z \mathrm{d} n. 
 \label{pieceD}
 \end{equation}
 The power constraint in (\ref{power_cons1}) is rewritten as
  \begin{equation}
 P =\sum\limits_{k=1}^{K} \int\limits _{\mathbb R} g_k^2(x)p_X(x)p_{K|X}(k|x) \mathrm{d} x.
 \label{power_new1}
 \end{equation}
 The cost function to minimize is
 \begin{align}
 J&=D+\lambda P \\
 &=\sum\limits_{k=1}^{K} \int\limits _{\mathbb R} J_k(x)p_{X}(x)p_{K|X}(k|x) \mathrm{d} x,
 \label{distortion_new1}
 \end{align}
 where 
 \begin{equation}
 J_k(x) \triangleq D_k(x)+\lambda g_k^2(x)  \quad \forall k.
 \label{pieceJ}
 \end{equation}
 
 We now restate the problem as that of minimizing $J$ over the local model parameters and association probabilities. Note that, given the local models, the association probabilities that minimize (\ref{distortion_new1}) will implement 'hard' associations, that is, every point is associated with probability one to the region that contributes the minimum cost to (\ref{pieceJ}). Therefore, by randomizing the encoder we generalize the search space but preserve the same global minimum as the original problem.
 
\subsection {Entropy Constraint}
As we noted above, the direct optimization of the association probabilities will result in 'hard' probabilities. However, in order to avoid poor local optima we impose and control the level of randomness, i.e. we introduce a constraint on the randomness of the encoder, which we measure by the Shannon entropy. The total entropy of the encoder is given by  $H(X,K) = H(X) + H(K|X)$ and since $H(X)$ is constant (determined by the source) we define $H \triangleq H(K|X) $ where
 \begin{equation}
 H(K|X) = -\int\limits _{\mathbb R}p_X (x)  \sum\limits_{k=1}^{K} p_{K|X}(k|x)\log(p_{K|X}(k|x))\mathrm{d} x.
 \end{equation}
 
 \begin{remark}
It is important to note that the approach is generalizable to the "mass-constrained" variant of DA \cite{rose1993constrained}, where entropy maximization is effectively replaced by minimization of the mutual information $I(K;X)$. Such generalization offers additional optimization advantages (see \cite{rose1993constrained}), as well as a useful and direct link to rate-distortion theory (see \cite{Roserd} for analysis of these connections, as well as DA for rate-distortion function computation). The corresponding "mass-constrained" extension for the current problem is a work in progress and is outside of the scope of this paper.
\end{remark}
 
 \begin{figure*}
\centering \includegraphics[scale=0.3]{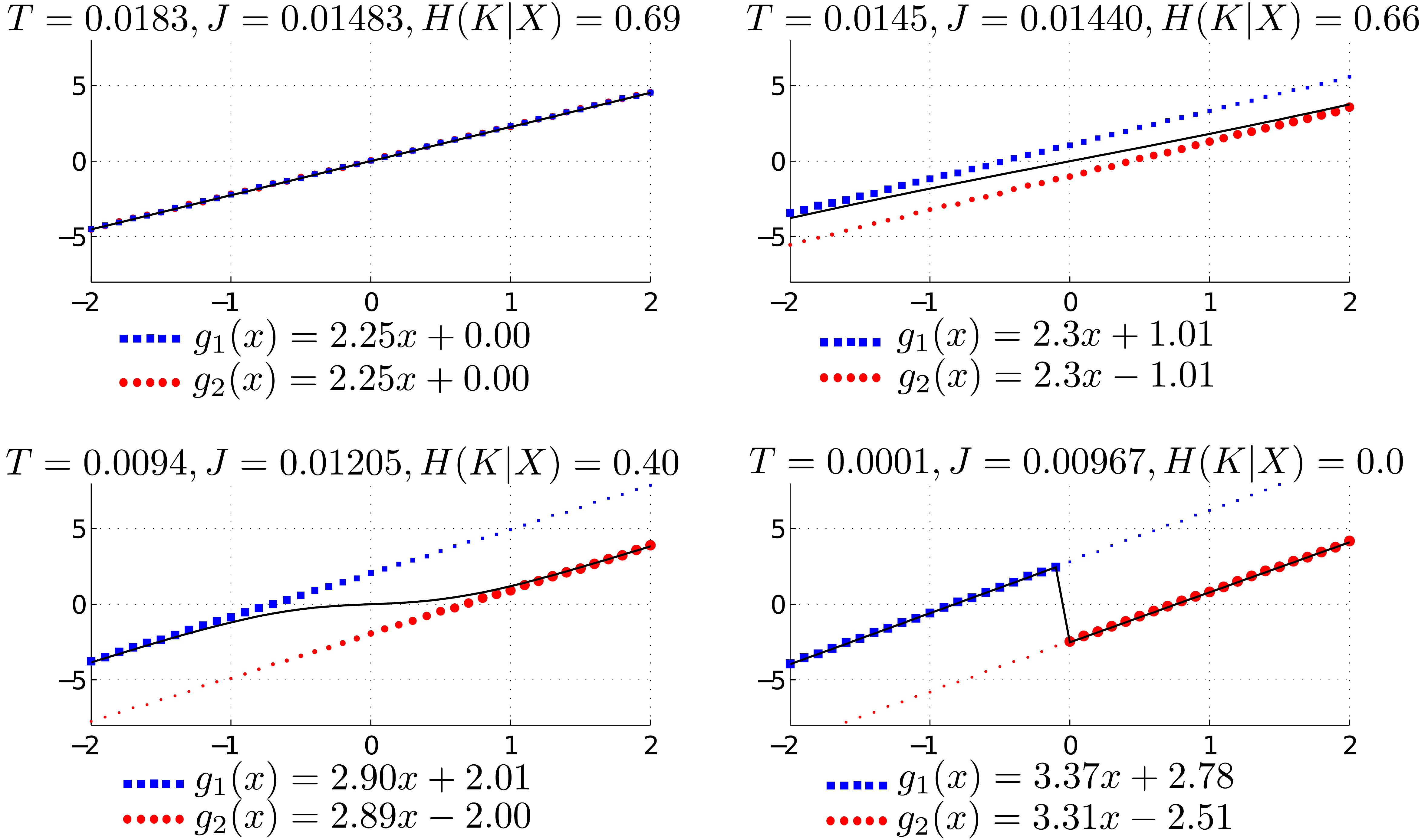}
\caption{The evolution of the encoder in the algorithm. The two models are shown by dotted lines and the size of a dot gives the probability association at that input point to that local model. The black line represents the averaged encoder, $K=2$.}
\label{evo1}
\end{figure*}

Accordingly we construct the Lagrangian
\begin{equation}
F=J-TH,
\label{a}
\end{equation}
to be minimized, with $T$ (temperature) being the Lagrange multiplier associated with the entropy constraint. Note that for large $T$, the minimum $F$ is achieved by maximizing the entropy. At lower values of $T$, randomness is traded for reduction in $J$. In the limit $T = 0$, minimizing $F$ corresponds to minimizing $J$ directly, which produces a deterministic encoder. Therefore,  we start at a high value of $T$ and gradually lower it while minimizing $F$ at each step.

We present an example of the method in Figure \ref{evo1} with two local models\footnote{The example is run for jointly Gaussian source and side information and Gaussian noise.}. When $T$ is large, the local models are coincident. As we lower $T$, the system goes through a bifurcation point  (referred as ``phase transition" in statistical physics) where the two local models split from each other to decrease $F$. The corresponding value of $T$ is referred as the first ``critical temperature". Further phase transitions can be obtained by keeping a duplicate for each local model at every temperature. The duplicates will merge  at every iteration until a critical temperature is reached, and will split at a phase transition.

The pseudocode of the method is given in Algorithm \ref{DAalgorithm1}.

\begin{algorithm}
\caption{ The outline of the proposed algorithm }\label{DAalgorithm1}
\begin{algorithmic}
\State Initialize: High T, single region (K=1)
\While{$H(K|X) > H_{min}(K|X)$ }
\State Duplicate (if $K<K_{max}$) and perturb local models
\While{$cost_{i+1}<cost_i$}
\State update the local model parameters 
\State update $p_{K|X}(k|x)$ $\forall k,x$
\State update $w(y,z)$ 
\EndWhile
\State Check if regions have split
\State Set $T=\alpha T$ \Comment{e.g. $\alpha = 0.95$}
\EndWhile
\end{algorithmic}
\end{algorithm}

\subsection{Update Equations} 
The optimum local model parameters cannot be obtained in closed form, hence we perform gradient descent search. The gradient with respect to any local parameter $\theta _k$ from a set $\Lambda _k$ can be obtained as
  \begin{equation}
\frac{\partial F}{\partial \theta _k}=  \frac{\partial J}{\partial \theta _k}= \int\limits_{x}p_X(x)p_{K|X}(k|x)\frac{\partial [D_k(x) +\lambda  g_k^2(x)]}{\partial \theta _k}\mathrm{d} x.
 \end{equation}  
For the affine model, $\theta_k$ denotes $a_k$ and $b_k$. 

The association probabilities that minimize $F$ are derived in a straightforward fashion as the Gibbs distribution
\begin{equation}
 p_{K|X}(k|x) = \frac{e^{-[D_k(x) + \lambda g_k^2(x)]/T} }{ \sum\limits_{k=1}^{K}e^{-[D_k(x) + \lambda g_k^2(x)]/T}} \quad \forall x.
 \label{optimum prob}
 \end{equation}

 \begin{remark}As expected, (\ref{optimum prob}) results in uniform associations for large $T$ and ``hard" (binary) associations for $T=0$.
 \end{remark}
 
 The optimum decoder given the encoder can be derived  by plugging $p(y|x,z) = \sum\limits_{k=1}^{K}p_N(y-g_k(x))p_{K|X}(k|x)$ in (\ref{decoder_map}).
 
\section {Experimental Results}

\begin{figure}
\centering
\includegraphics[scale=0.55]{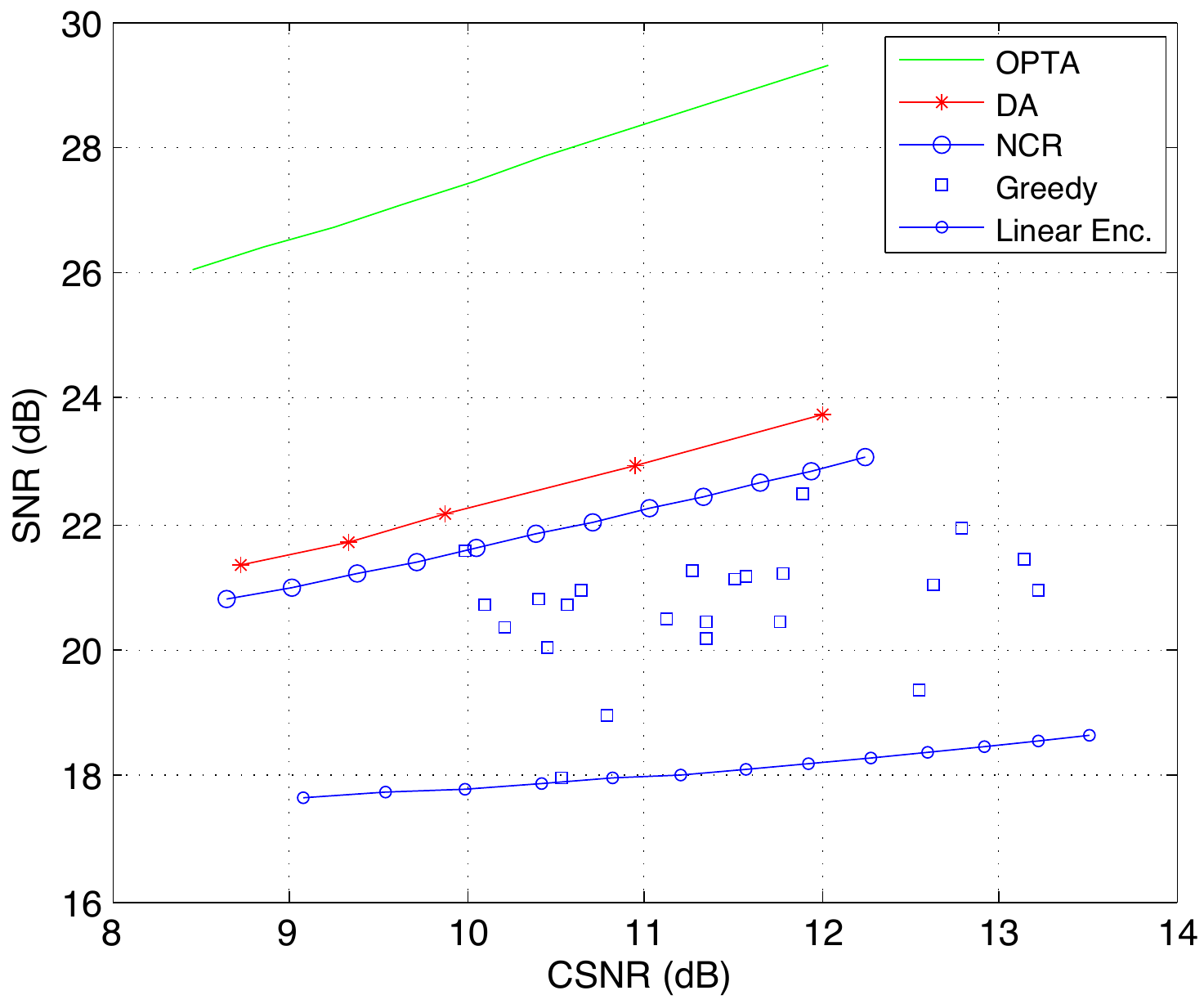}
\caption{The performance comparison of the proposed method with greedy optimization, the noisy relaxation (NCR), and the linear mappings.}
\label{compSide} 
\end{figure}

\begin{figure}
\includegraphics[width=1 \linewidth]{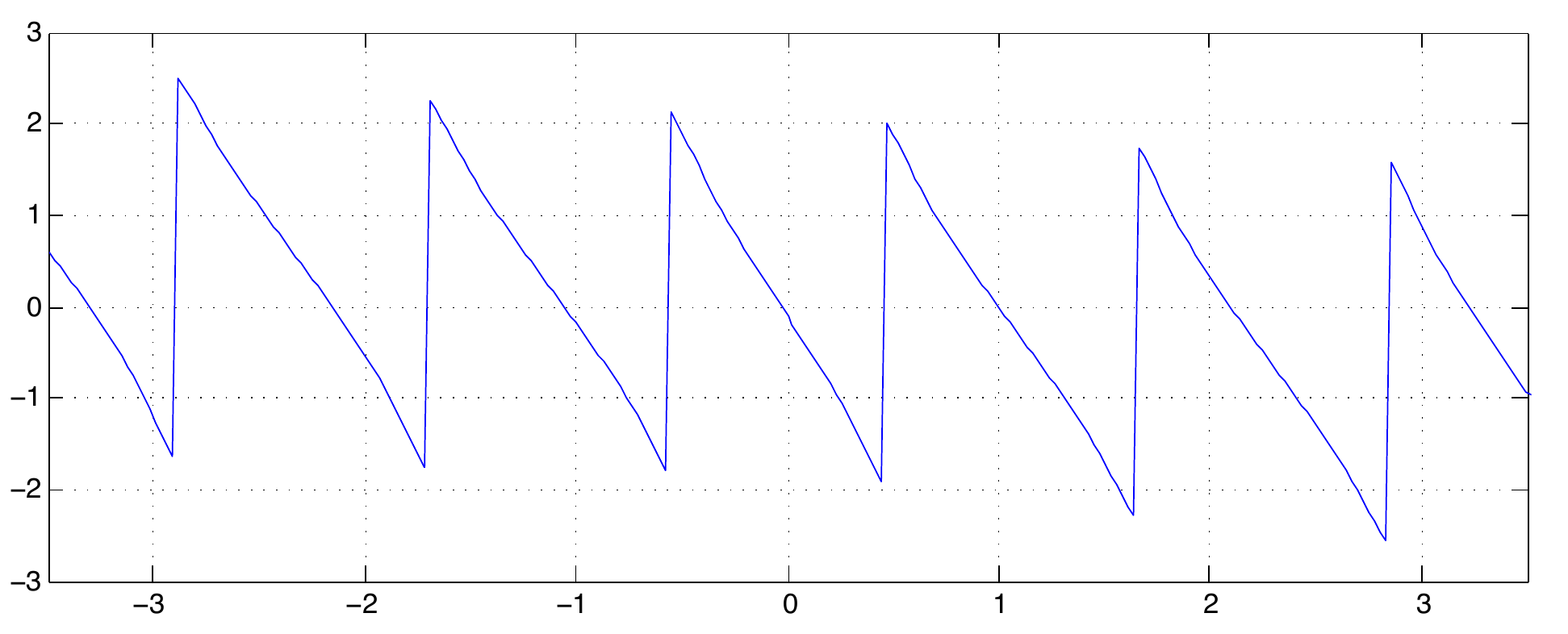}
\caption{An example mapping for correlation coefficient $\rho=0.99$ at $CSNR = 10.98$, $SNR=23.2$.}
\label{sideEx1} 
\end{figure}

The comparative performance results are given in Figure \ref{compSide} for jointly Gaussian $X$ and $Z$  with $\rho=0.99$ as described in Section II, and Gaussian noise with unit variance. The best results of the NCR method out of multiple runs, and various results from the greedy method are presented in Figure \ref{compSide}. Note that the proposed method is independent of the initialization and only run once, whereas the results of greedy approach and NCR heavily depend on initialization, as can be seen from various points obtained by the greedy approach. We also present the performance of OPTA as benchmark while noting that it is asymptotic and requires infinite delay. The performance of linear encoder and decoder is plotted as well, since it is also a local minimum (see Remark \ref{linear necessary}). 

An example mapping from the same setting is also given in Figure \ref{sideEx1}. Interestingly,  as noted before (see e.g., \cite{emrah_dcc10,karlsson2010optimized }) the analog mapping captures the central characteristic observed in digital Wyner-Ziv mappings, in the sense of many-to-one mappings, where multiple source intervals are mapped to the same channel interval, which will potentially be resolved by the decoder given the side information. However, we see  differences between the mappings obtained by NCR (see e.g.,\cite{emrah_dcc10,karlsson2010optimized }) and ones by the proposed DA based method,  e.g,  the linear trend of the encoding mapping, that yield significant performance improvement as shown in Figure  \ref{compSide}. Such differences are difficult to obtain and very important for the design of parametric mappings, see e.g., \cite{chen2011zero}.

\section {Conclusions}
In this paper, we studied the problem of finding globally optimum encoder and decoder pairs in zero delay source-channel coding, focusing on the setting where a side information is available to the decoder. Since the cost surface is riddled with locally optimum points, we developed a method based on the deterministic annealing approach to obtain globally optimum points. The numerical results show superiority of the proposed algorithm over greedy optimization methods and as well as the previously adopted approach, i.e., NCR. As future work, we will investigate adopting our DA approach to obtain optimal mappings in more complicated network settings as well as well-known open control problems such as the Witsenhausen's counterexample. 

\bibliographystyle{IEEEtran}
\bibliography{ref}
\end{document}